\documentclass[10pt,twocolumn,letterpaper]{article}

\usepackage{iccv}
\usepackage{times}
\usepackage{graphicx}
\usepackage{amsmath}
\usepackage{amssymb}
\usepackage{subfig}
\usepackage{multirow}

\makeatletter
\renewcommand{\maketag@@@}[1]{\hbox{\m@th\normalsize\normalfont#1}}%
\makeatother

\usepackage[pagebackref=true,breaklinks=true,letterpaper=true,colorlinks,bookmarks=false]{hyperref}

\iccvfinalcopy 

\ificcvfinal\fi

\begin{document}

\title{Rational Polynomial Camera Model Warping for Deep Learning Based Satellite Multi-View Stereo Matching}

\author{Jian Gao\footnotemark[1]\qquad Jin Liu\footnotemark[1]\qquad Shunping Ji\footnotemark[2]\\
School of Remote Sensing and Information Engineering, Wuhan University, China\\
{\tt\small \{jian\_gao,liujinwhu,jishunping\}@whu.edu.cn}\\
}

\maketitle

\renewcommand{\thefootnote}{\fnsymbol{footnote}}
\footnotetext[1]{Equally contributed.}
\footnotetext[2]{Corresponding author.}

\ificcvfinal\thispagestyle{empty}\fi

\begin{abstract}
Satellite multi-view stereo (MVS) imagery is particularly suited for large-scale Earth surface reconstruction. Differing from the perspective camera model (pin-hole model) that is commonly used for close-range and aerial cameras, the cubic rational polynomial camera   (RPC) model is the mainstream model for push-broom linear-array satellite cameras. However, the homography warping used in the prevailing learning based MVS methods is only applicable to pin-hole cameras. In order to apply the SOTA learning based MVS technology to the satellite MVS task for large-scale Earth surface reconstruction, RPC warping should be considered. In this work, we propose, for the first time, a rigorous RPC warping module. The rational polynomial coefficients are recorded as a tensor, and the RPC warping is formulated as a series of tensor transformations. Based on the RPC warping, we propose the deep learning based satellite MVS (SatMVS) framework for large-scale and wide depth range Earth surface reconstruction. We also introduce a large-scale satellite image dataset consisting of 519 5120$\times$5120 images, which we call the TLC SatMVS dataset. The satellite images were acquired from a three-line camera (TLC) that catches triple-view images simultaneously, forming a valuable supplement to the existing open-source WorldView-3 datasets with single-scanline images. Experiments show that the proposed RPC warping module and the SatMVS framework can achieve a superior reconstruction accuracy compared to the pin-hole fitting method and conventional MVS methods. Code and data are available at \href{https://github.com/WHU-GPCV/SatMVS}{ https://github.com/WHU-GPCV/SatMVS}.
\end{abstract}

\section{Introduction}
Three-dimensional (3D) scene reconstruction from multi-view optical images is a hot topic in both the computer vision and remote sensing fields. Currently, there are two mainstream camera models: the perspective camera model (i.e., the pin-hole model) and the cubic rational polynomial camera (RPC) model. The prevailing deep learning based multi-view stereo (MVS) methods \cite{9115828, Cheng2020DeepSU,Gu2020CascadeCV,Liu2020ANR,Luo2019PMVSNetLP,Yang2020CostVP, Yao2018MVSNetDI,Yao2019RecurrentMF} have been designed and developed for multi-view images captured from pin-hole cameras. To date, the deep learning based MVS task for RPC-based images, i.e., optical satellite images, has not been tackled. \\
The study of satellite MVS task is very important as the multi-view satellite images captured from push-broom linear-array cameras with the RPC model are the main data source for large-scale 3D Earth surface reconstruction. In the recent deep learning based MVS methods \cite{Yao2018MVSNetDI, Cheng2020DeepSU, Gu2020CascadeCV}, homography warping has been utilized to align images from different viewpoints to a reference view through a set of hypothetical fronto-parallel depth planes of the reference camera \cite{Yao2018MVSNetDI}. Analogously, developing a specific warping method for aligning multi-view satellite images to the reference image space could extend the deep learning based MVS methods to satellite images. The primary contribution of this work is that we propose, for the first time, a rigorous RPC warping module. Based on the RPC warping module, we then propose the SatMVS framework for deep learning based satellite MVS task. The SatMVS framework utilizes RPC warping to align the multi-view satellite images through the hypothetical height planes in the world coordinate system instead of the front-parallel depth planes. It features a multi-stage coarse-to-fine matching strategy on a feature pyramid to tolerate the wide search range, considering the great topographic relief of the Earth’s surface and high-capacity satellite images. The SatMVS framework can also be used with most of the modern deep learning based MVS methods by replacing the perspective model with the RPC model and introducing the multi-stage strategy. \\
The other contribution is that we introduce a new open-source satellite MVS dataset (the TLC SatMVS dataset). We believe that this is the first open-source satellite dataset to be built from images collected by the push-broom three-line camera (TLC). The camera is mounted on the Ziyuan-3 (ZY-3) satellite. This dataset will be a beneficial complement to the existing single linear-array WorldView-3 datasets such as MVS3D \cite{Bosch2016AMV} and US3D \cite{Bosch2019SemanticSF}. The difference between a TLC and a single-scanline camera is that the former captures three images at the same scene simultaneously, and the latter has to shoot from different orbit positions at different times to form a stereo picture of the same scene.\\ 
In summary, we fill two gaps in this paper: 1) the lack of RPC warping module, which has hindered the application of the state-of-the-art MVS method to satellite images; and 2) the lack of large open-source TLC satellite MVS datasets, which has hampered the development of satellite MVS and large-scale Earth surface reconstruction.
\section{Related Work}
\textbf{Datasets.} Most public stereo datasets have been created for the reconstruction of close-range or small-scale scenes. Examples of the two-view stereo benchmark datasets are the Middlebury \cite{Scharstein2004ATA}, KITTI\cite{Geiger2012AreWR}, and Sceneflow\cite{Mayer2016ALD} datasets. MVS benchmark datasets include the Middlebury-MVS\cite{Seitz2006ACA}, DTU\cite{Aans2016LargeScaleDF}, Tanks and Temples\cite{Knapitsch2017TanksAT}, and ETH3D\cite{Schps2017AMS} datasets. As a supplement to these close-range datasets, Liu and Ji\cite{Liu2020ANR} created the WHU MVS/Stereo dataset, which is the first large-scale multi-view aerial image dataset for the task of city-level Earth surface reconstruction. Multi-view satellite images are another important data source for Earth surface reconstruction. Currently, there are two public multi-view satellite datasets, both of which consist of WorldView-3 single linear array images. One is the MVS3D dataset\cite{Bosch2016AMV}, which provides 47 panchromatic images covering an area of approximately 100 $km^{2}$ near San Fernando, Argentina. However, the airborne LiDAR ground-truth data only cover a sub-area of about 20 $km^{2}$, which is insufficient for the deep learning based MVS methods. The other is the US3D dataset\cite{Bosch2019SemanticSF}, which is made up of 69 WorldView-3 images and covers a total area of 100 $km^{2}$ of two cities in the United States. The WorldView-3 single linear-array camera acquires MVS images on different dates. As a result, the temporal and seasonal changes in the MVS3D and US3D datasets cause obvious visual differences among the MVS images, leading to a negative effect on the MVS task.\\
\textbf{3D Surface Reconstruction with Satellite Images.} 3D Earth surface reconstruction from satellite imagery is mainly achieved through the traditional geometric methods, which can be roughly divided into two main types. The first type of method is based on the epipolar geometry of satellite images. An example of this is the RPC Stereo Processor (RSP)\cite{isprs-annals-III-1-77-2016}. In this type, the stereo images are first rectified according to the RPC model\cite{Wang2011EpipolarRO}, and a stereo matching algorithm such as the semi-global matching (SGM) stereo method \cite{Hirschmller2008StereoPB} is then used to estimate the disparities. Finally, the disparity maps are converted into 3D points in the world coordinate system. The other type involves fitting a complex RPC model into a pin-hole model in a small area, and then using the stereo/MVS pipeline for the reconstruction. Examples of this type are the satellite stereo pipeline (S2P) \cite{Franchis2014AnAA} for binocular stereo imagery and adapted COLMAP \cite{Zhang2019LeveragingVR} for MVS imagery.\\
\textbf{Cost Volume Construction.} Cost volume construction is an essential part of the learning-based stereo/MVS methods. GC-Net \cite{Kendall2017EndtoEndLO} generates a 3D cost volume in stereo matching by concatenating the left and right feature maps along the disparity direction. \cite{Hartmann2017LearnedMS} introduced a learned cost metric for MVS task by considering multi-patch similarity. SurfaceNet \cite{Ji2017SurfaceNetAE} and DeepMVS \cite{Huang2018DeepMVSLM} pre-warp the multi-view images to the sweep planes in 3D space. Most of the recent state-of-the-art methods \cite{9115828, Cheng2020DeepSU,Gu2020CascadeCV,Liu2020ANR,Luo2019PMVSNetLP,Yang2020CostVP, Yao2018MVSNetDI,Yao2019RecurrentMF} use differentiable homography warping to construct the cost volume or cost maps. Given a set of fronto-parallel planes of the reference camera at different depths, the images (or the extracted features) of each view are warped into the view of reference camera by a 3$\times$3 homography matrix, and then the warped feature volumes are fused to a single cost volume. However, the homography matrix is derived from the pin-hole model, so that it is not suitable for multi-view satellite images with the RPC model. \cite{Zhang2019LeveragingVR} proposed a solution by fitting the RPC model into a pin-hole model in small patches of a satellite image. However, this fitting strategy breaks the strict geometric relationship of the imaging model, and introduces inevitable errors, while also requiring a massive amount of preprocessing.\\
\textbf{Deep Learning Based MVS Methods.} The deep learning based MVS approaches can be classified into: 1) 3D convolution based methods, such as MVSNet \cite{Yao2018MVSNetDI}, MVSNet++ \cite{9115828}, and P-MVSNet \cite{Luo2019PMVSNetLP}, which apply a series of 3D convolutions to regularize the cost volume; and 2) recurrent regularization based methods, such as R-MVSNet \cite{Yao2019RecurrentMF}, which depth-wise process the cost maps of different depths. The former methods are intuitive, but require much more GPU capacity. All of these methods were developed for use with natural images. RED-Net \cite{Liu2020ANR} extends the convolutional gated recurrent unit (ConvGRU) based regularization method \cite{Yao2019RecurrentMF} for aerial MVS task. Methods based on building cost volumes at multiple stages have also been recently introduced, e.g., CasMVSNet \cite{Yao2018MVSNetDI}, CVP-MVSNet \cite{Yang2020CostVP}, and UCS-Net\cite{Cheng2020DeepSU}, in which a coarse-to-fine pyramid matching structure is applied. These multi-stage methods narrow the depth search range of the current stage to construct a thin cost volume with lower memory and a higher depth-wise sampling rate, which is conducive to large-scale scene reconstruction. However, it is difficult to apply the state-of-the-art deep learning based MVS methods to satellite imagery with a complex RPC model. In this paper, we attempt to fill this gap by proposing a general deep learning based MVS framework for satellite images (SatMVS) with a novel rigorous RPC warping module.
\begin{figure*}
\begin{center}
\subfloat[]{
\includegraphics[width=0.23\linewidth]{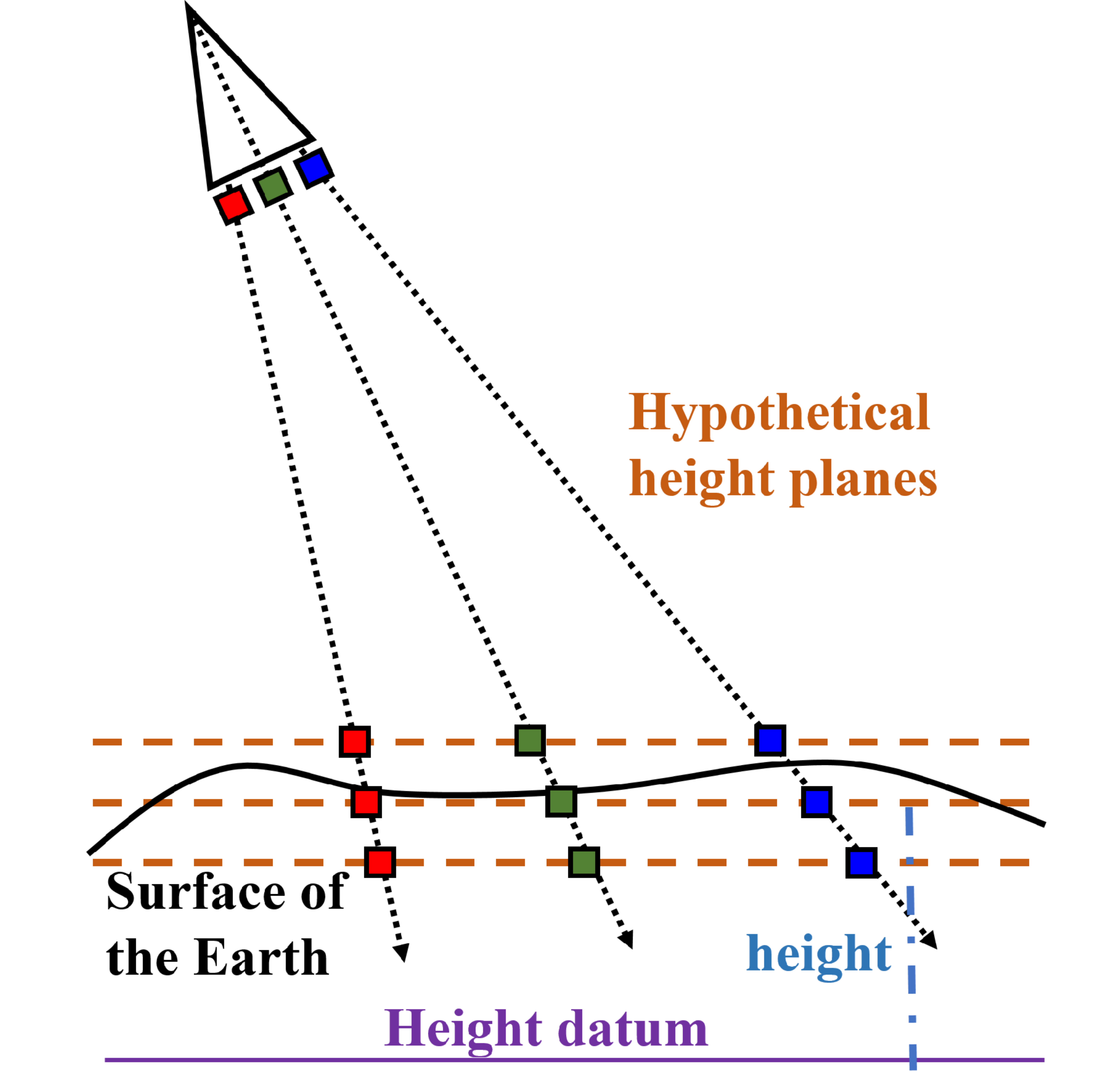}}
\setlength{\abovecaptionskip}{-0.0cm}
\label{1(a)}
\vspace{-0.0cm}
\subfloat[]{
\includegraphics[width=0.23\linewidth]{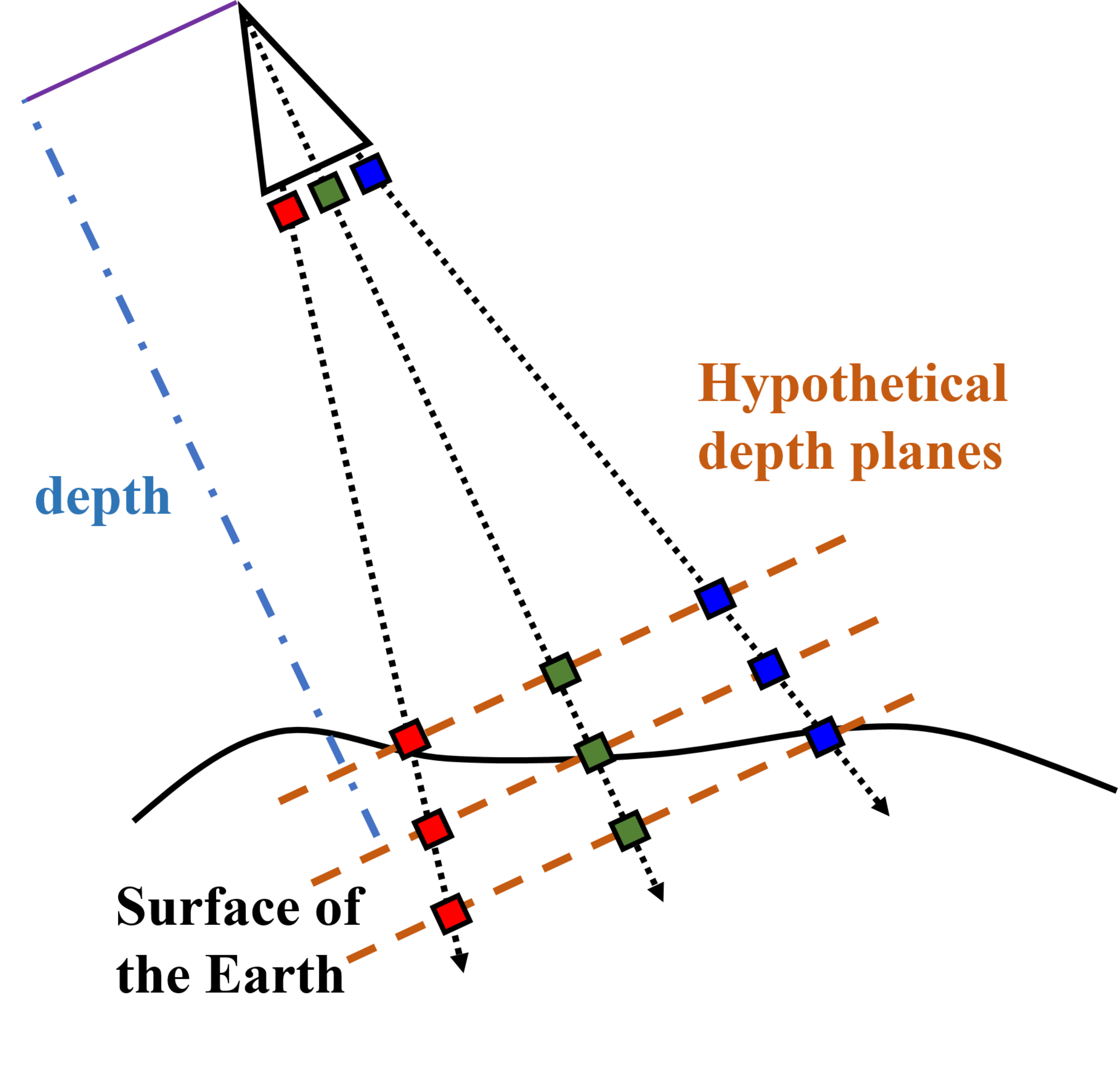}}
\setlength{\abovecaptionskip}{-0.0cm}
\label{1(b)}
\vspace{-0.0cm}
\subfloat[]{
\includegraphics[width=0.23\linewidth]{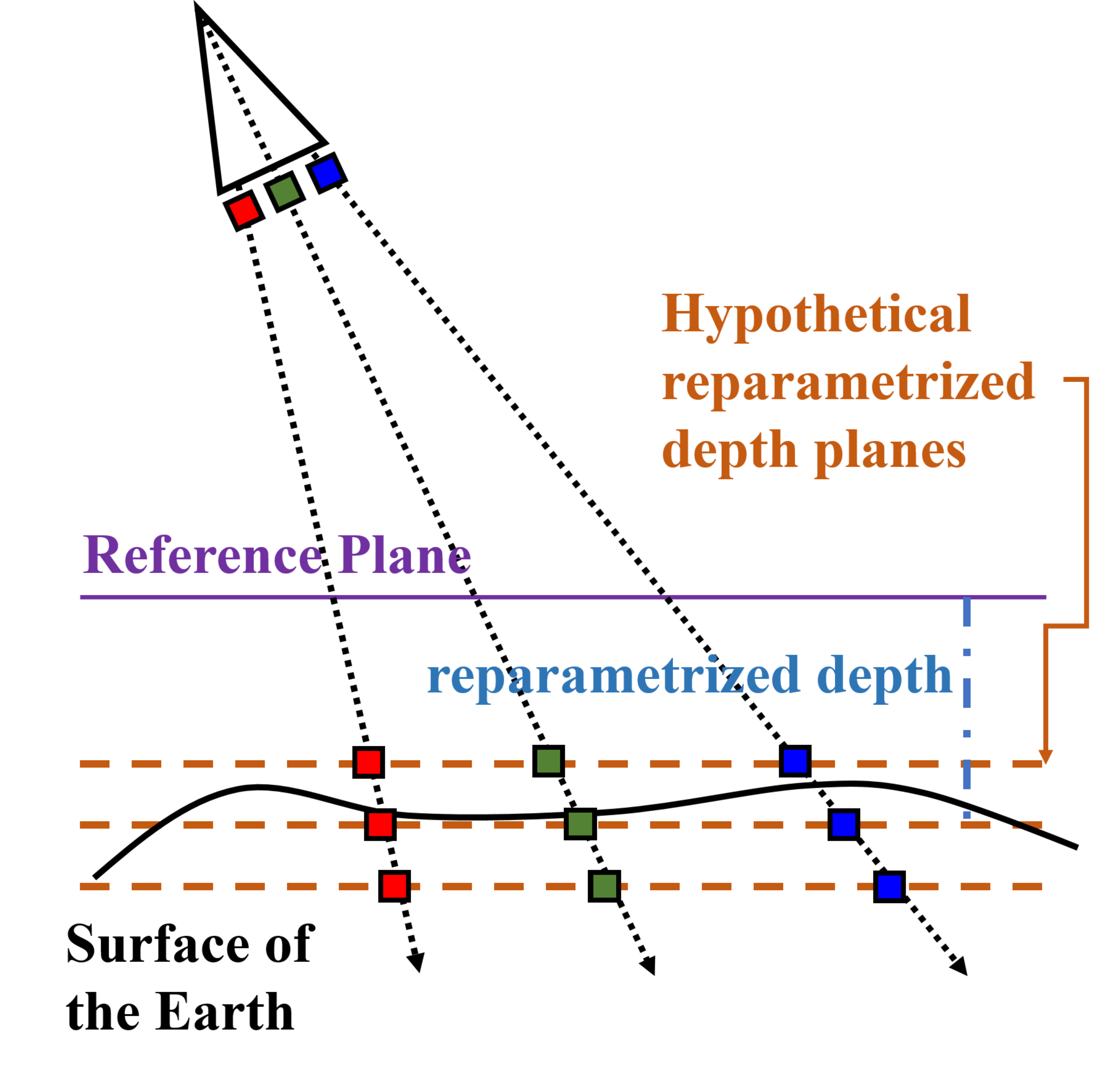}}
\setlength{\abovecaptionskip}{-0.0cm}
\label{1(c)}
\end{center}
\setlength{\abovecaptionskip}{-0.3cm}
   \caption{The hypothetical (a) height planes in our RPC warping, (b) depth planes, and (c) reparametrized depth planes.}
\label{fig:1}
\vspace{-0.3cm}
\end{figure*}
\section{RPC Warping}
\subsection{Rational Polynomial Camera Model}
The RPC model is the most widely used camera model in high-resolution satellite images. It connects the image points and corresponding world coordinate points with cubic rational polynomial coefficients. In Eq.(1), a 3D point in the normalized world coordinates $(latitude, longitude, height)$, denoted as $(lat_{n}, lon_{n}, hei_{n})$, is transformed to image space to obtain the normalized image coordinates $(samp_{n}, line_{n})$, which correspond to the along-array and along-track directions, respectively. Eq.(2) is the reverse version. $P^{fwd}$ and $P^{inv}$ are both cubic polynomials, as shown in Eq.(3), where the summation of integers $m_{1}$, $m_{2}$ and $m_{3}$ is no more than 3.
\begin{equation}
\left\{\begin{array}{l}
{samp}_{n}=\frac{P_{1}^{fwd}\left({lat}_{n},{lon}_{n},{hei}_{n}\right)}{P_{2}^{fwd}\left({lat}_{n},{lon}_{n},{hei}_{n}\right)} \\
{line}_{n}=\frac{P_{3}^{fwd}\left({lat}_{n},{lon}_{n},{hei}_{n}\right)}{P_{4}^{fwd }\left({lat}_{n},{lon}_{n},{hei}_{n}\right)}
\end{array}\right.
\end{equation}
\begin{equation}
\left\{\begin{array}{l}
{lat}_{n}=\frac{P_{1}^{inv}\left(samp_{n}, line_{n},hei_{n}\right)}{P_{2}^{inv}\left(samp_{n},line_{n}, hei_{n}\right)} \\
{lon}_{n}=\frac{P_{3}^{inv}\left(samp_{n}, line_{n},hei_{n}\right)}{P_{4}^{inv}\left(samp_{n}, line_{n}, hei_{n}\right)}
\end{array}\right.
\end{equation}
\begin{equation}
P(X, Y, Z)=\sum_{i=0}^{m_{1}} \sum_{j=0}^{m_{2}} \sum_{k=0}^{m_{3}} c_{i j k} \cdot X^{i} \cdot Y^{i} \cdot Z^{i}
\end{equation}
Clearly, the RPC model is a general geometric model, instead of a physical camera model; however, it has been proved that the RPC model can achieve a very high accuracy comparable to the rigorous sensor model (RSM) \cite{Tao2001ACS, Grodecki2003BlockAO}, guaranteeing its successful application to all of the high-resolution optical satellite images.
\begin{figure}
\begin{center}
\includegraphics[width=0.45\linewidth]{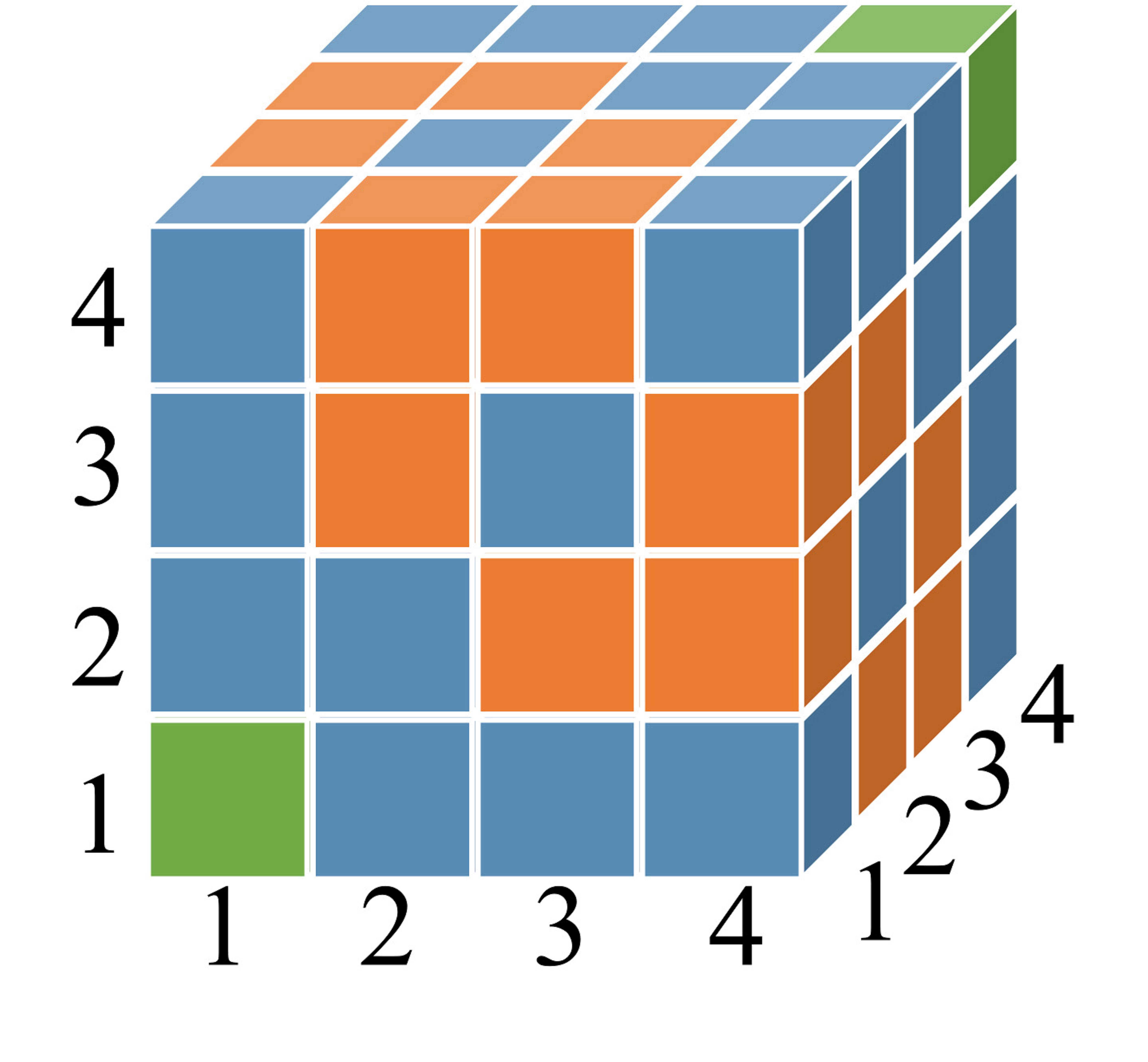}
\end{center}
\setlength{\abovecaptionskip}{-0.3cm}
   \caption{Coefficient tensor $\boldsymbol T$. When $i$, $j$, and $k$ are equal, $\boldsymbol T_{ijk} = a_{i}a_{j}a_{k}$(green blocks); when there are only two of $i$, $j$, and $k$ equal, $\boldsymbol T_{ijk}= a_{i}a_{j}a_{k}/ 3$ (blue blocks); when none of $i$, $j$, and $k$ are equal, $\boldsymbol T_{ijk}=a_{i}a_{j}a_{k}/ 6$ (orange blocks). }
\label{fig:2}
\vspace{-0.7cm}
\end{figure}
\begin{figure*}
\begin{center}
\includegraphics[width=0.9\linewidth]{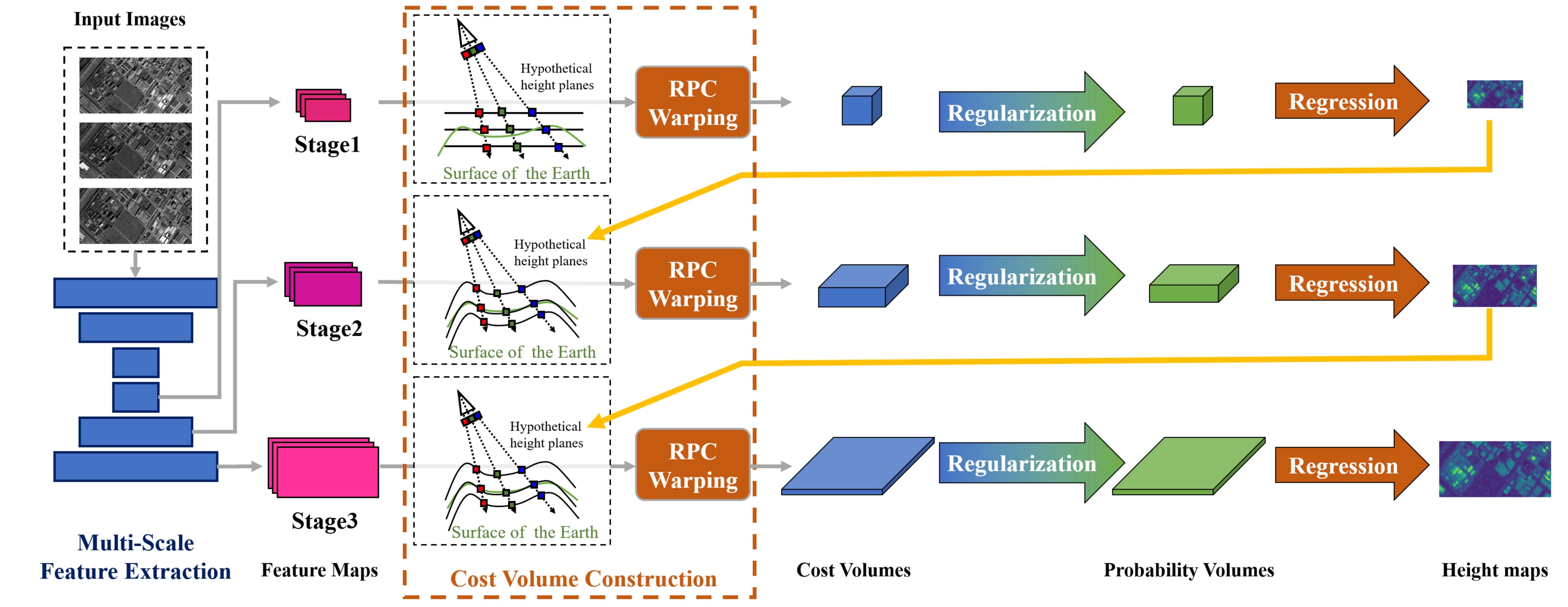}
\end{center}
\setlength{\abovecaptionskip}{-0.2cm}
   \caption{The structure of the SatMVS framework.}
\label{fig:3}
\vspace{-0.4cm}
\end{figure*}
\subsection{Differentiable RPC Warping}
In the pin-hole camera model, the geometric transformation between the corresponding pixels of stereo images can be formulated as a simple $3\times3$ homography matrix by given depths. Almost all of the state-of-the-art MVS methods \cite{9115828, Cheng2020DeepSU,Gu2020CascadeCV,Liu2020ANR,Luo2019PMVSNetLP,Yang2020CostVP, Yao2018MVSNetDI,Yao2019RecurrentMF} warp the source views to the reference view through a homography matrix and a series of hypothetical fronto-parallel planes of the reference view. In contrast, the RPC model is much more complicated and its warping cannot be formulated with only a matrix. In this paper, we propose a rigorous and efficient RPC warping module, which is fundamentally a high-order tensor transformation. It projects images (or extracted features) from different views to the reference view through a set of hypothetical height planes in the world coordinate system, instead of the fronto-parallel planes of a reference view, because there are no explicit physical parameters in RPC model to define what is the front of a camera. Fig. \ref{fig:1} shows the difference between the hypothetical height planes we used, the hypothetical depth planes, and reparametrized depth planes used in \cite{Zhang2019LeveragingVR}.\\
We start by transforming the ternary cubic polynomial in Eq.(3) to a quaternion cubic homogeneous polynomial $f\left(x_{1}, x_{2}, x_{3},x_{4}\right)=\sum\left(a_{i} a_{j} a_{k}\right) \cdot x_{i} x_{i} x_{k}$, where $x_{1}{\equiv}1$ and $a_{i}a_{j}a_{k}(i,j,k\in{\{1, 2, 3, 4\}})$. By setting $x_{2}=lon_{n}$, $x_{3}=lat_{n}$, and $x_{4}=hei_{n}$, $f$ becomes the numerator or denominator of the RPC forward form, and that of the inverse form by setting $x_{2} = line_{n}$, $x_{3} = samp_{n}$, $x_{4} = hei_{n}$. Please note that we use Einstein’s summation convention in all of the formulas in this section. The four variables $x_{1}$, $x_{2}$, $x_{3}$, $x_{4}$ are then expressed as a tensor $\boldsymbol X$  with a rank of 1: $\boldsymbol X = (x_{1}, x_{2}, x_{3}, x_{4})^{T}$, and the polynomial coefficients are expressed as a tensor $\boldsymbol T$ with a rank of 3 and a shape of $4\times4\times4$ (see Fig. \ref{fig:2}). Through the tensor contraction operation, which calculates the sum over all the possible values of the repeated indices of a set of tensors \cite{Orus2013API}, the numerator and denominator of the RPC model can be expressed in tensor form:  
\begin{equation}
f(\boldsymbol{X})= \boldsymbol{T}_{i j k} \boldsymbol{X}_{i} \boldsymbol{X}_{j} \boldsymbol{X}_{k}
\end{equation}
The relationship between the elements in $\boldsymbol T_{ijk}$ and $a_{i}a_{j}a_{k}$ is: when $i$, $j$, and $k$ are equal, $\boldsymbol T_{ijk} = a_{i}a_{j}a_{k}$; when there are only two of $i$, $j$, and $k$ equal, $\boldsymbol T_{ijk}= a_{i}a_{j}a_{k}/ 3$; when none of $i$, $j$, and $k$ are equal, $\boldsymbol T_{ijk}=a_{i}a_{j}a_{k}/ 6$. Finally, the left side of Eq.(1) or (2) can be easily obtained by dividing the numerator and denominator in an element-wise way. We can then extend this calculation to adapt to both the batch operation and the RPC transformation with a set of points:
\begin{equation}
f^{(b m)}(\boldsymbol{X})=\boldsymbol{T}_{i j k}^{(b)} \boldsymbol{X}_{i}^{(b m)} \boldsymbol{X}_{j}^{(b m)} \boldsymbol{X}_{k}^{(b m)}
\end{equation}
where $X^{(bm)}$ represents the $m$-th point in the $b$-th batch and $T^{(b)}$ represents the coefficient tensor in the $b$-th batch. The elements in parentheses do not participate in the summation. Through element-wise division, the RPC warping of all the points in a batch can be calculated in one shot.\\
Specifically, for a point $(samp_{s}, line_{s})$ in the source image and a given plane $hei$, we obtain the normalization form $(samp_{s,n}, line_{s,n}, hei_{n})$ according to the available normalization parameters, which are also a part of the RPC parameters, and construct tensor  $\boldsymbol {X}_{s}=(1, samp_{s,n}, line_{s,n}, hei_{n})$, and then warp it onto the reference view through:
\vspace{-0.1cm}
\begin{equation}
\left\{\begin{array}{l}
{ samp }_{r, n}=\frac{f_{1}^{fwd}\left(\boldsymbol{X}_{0}\right)}{f_{2}^{fwd}\left(\boldsymbol{X}_{0}\right)} \\
{ line }_{r, n}=\frac{f_{3}^{fwd}\left(\boldsymbol{X}_{0}\right)}{f_{4}^{fwd}\left(\boldsymbol{X}_{0}\right)}
\end{array}\right.
\end{equation}
where $\boldsymbol{X}_{0}=\left(1, \frac{f_{1}^{{inv }}\left(\boldsymbol{X}_{s}\right)}{f_{2}^{inv }\left(\boldsymbol{X}_{s}\right)}, \frac{f_{3}^ {inv }\left(\boldsymbol{X}_{s}\right)}{f_{4}^ {inv }\left(\boldsymbol{X}_{s}\right)}, h e i_{n}\right)$, and $\frac{(\cdot)}{(\cdot)}$ represents element-wise division.\\
After anti-normalization, the point $(samp_{s}, line_{s})$ is warped to the corresponding point $(samp_{r}, line_{r})$ with differentiable resampling, e.g., bilinear interpolation, to complete the RPC warping.
\section{The Learning Based SatMVS Framework}
We propose a satellite MVS deep learning framework imbedded with RPC warping, which we call the SatMVS framework. In addition to the necessary part of RPC warping, we consider that coarse-to-fine multi-stage processing is also necessary in such a framework to predict a wide range of Earth surface elevation. The other parts, such as the feature extraction, cost map regularization, and regression module, can be borrowed from the state-of-the-art methods \cite{Gu2020CascadeCV,Liu2020ANR,Yang2020CostVP}. Finally, we obtain the complete SatMVS framework, as shown in Fig. \ref{fig:3}.
\subsection{Multi-scale Feature Extraction}
The current multi-scale MVS methods utilize popular feature extractors such as a feature pyramid network \cite{Lin2017FeaturePN} or UNet \cite{Ronneberger2015UNetCN}. All of these can be used in the SatMVS framework. For the experiments, we adopted a weight-shared multi-scale feature extractor \cite{Cheng2020DeepSU}, which consists of an encoder and a decoder with skip connections, to extract the features. The module outputs a three-scale feature pyramid with the size of \{1/16, 1/4 ,1\} of the input image size, and the numbers of channels are 32, 16, and 8, respectively.
\subsection{Multi-Stage Cost Volume Construction with RPC Warping}
There may be huge elevation differences in the hundreds of kilometers of landscape covered by a satellite image. We suppose that the maximum height difference is 2$km$. If the height interval is set to 2.5$m$, which is roughly the pixel resolution of ZY-3 imagery, 800 hypothetical planes can cover the entire elevation range with a pixel-level accuracy. In contrast, the depth search range of the close-range datasets is relatively small (e.g., 128 or 256 planes in the DTU dataset). To reduce the high demand for GPU memory and increase the learning speed, multi-scale learning from the feature pyramid is applied. \\
At the first stage ($i=1$), the search scope should cover the range between the maximum and minimum heights of the covered area. This information can be found in the available RPC parameters. Alternatively, several open-source global digital elevation model (DEM) products, such as the Shuttle Radar Topography Mission (SRTM) DEM \cite{Jarvis2008HolefilledSF} or the Advanced Spaceborne Thermal Emission and Reflection Radiometer (ASTER) global digital elevation model (GDEM) \cite{DTED2009ASTERGD} can be used as information sources. The number of hypothetical planes is fixed in the first stage (e.g., 64). The height interval is then determined by dividing the height difference by the number of hypothetical planes.\\ 
For the stage $i$ \!($i \!\geqslant \!2$), the interval and the number of hypothetical planes are empirically fixed, and the hypothetical planes are centered at the reference height of the last stage.\\
The feature maps from the source views are warped to the reference view through RPC warping to form multiple feature volumes, which are then fused to the cost volume using a variance-based operation \cite{Yao2018MVSNetDI}.
\subsection{Regularization}
The recent MVS networks use either a series of 3D convolutions \cite{Yao2018MVSNetDI} or 2D convolutional GRUs \cite{Yao2019RecurrentMF} to regularize the cost volume. Both can be used in the SatMVS framework. In practice, we use the recurrent encoder-decoder structure of RED-Net \cite{Liu2020ANR} to regularize the cost maps that are constructed at each stage. Compared to 3D convolutions, the RED structure sequentially regularizes the cost maps along the elevation direction to achieve a high efficiency and low memory cost, which significantly benefits the processing of high-capacity satellite images with a wide height search range.
\subsection{Height Inference}
After regularization, a \emph{soft argmin} operation is applied along the height direction for sub-pixel estimation. At the training stage, the pyramid network outputs height maps at three resolutions. Similar to the current works \cite{Bosch2019SemanticSF, Gu2020CascadeCV}, the total loss is defined as the weighted sum of the three-stage L1 loss, in which the weights are \{0.5, 1, 2\}, respectively.
\subsection{Pipeline}
We designed a complete pipeline for the RPC-based networks to reconstruct the final DSM, including 1) image preprocessing; 2) MVS inference; and 3) DSM generation.\\ 
In the preprocessing, the study area in the world coordinate system (usually the WGS-84 coordinate system) is divided into regular blocks, and each block is projected onto the MVS satellite images with the maximum and minimum elevation planes in this block respectively. The minimum bounding rectangles of the two projection regions in different views are calculated, extended to a uniform size, and cropped as the input of the network. This process ensures sufficient overlap between the cropped image patches of multiple views. In addition, gamma correction and linear enhancement can be used to improve the contrast. \\
In the MVS inference, the cropped images are fed into the proposed SatMVS framework. Each view image is treated as the reference image to infer the height map in turn. \\
In the DSM generation, the geometric consistency \cite{Yao2018MVSNetDI} is used to filter out outliers in the inferred height maps of the different views. If the distance between the reprojected point from the $j$-th view and the original point in the reference view is less than 1 pixel, the estimated result is considered to be geometrically consistent and valid. The valid matching results are then transformed to the Universal Transverse Mercator (UTM) coordinate system and resampled to a regularized DSM.\\
We also propose the homography warping version of the SatMVS pipeline. For the homography warping based networks, which cannot be directly used for satellite imagery, the RPC model of each cropped image is fitted to the pin-hole model according to \cite{Zhang2019LeveragingVR} in the image preprocessing part. In the parts of MVS inference and DSM generation, the fitting model is consistently used instead of the rigorous RPC model.
\begin{figure}
\begin{center}
\includegraphics[width=0.8\linewidth]{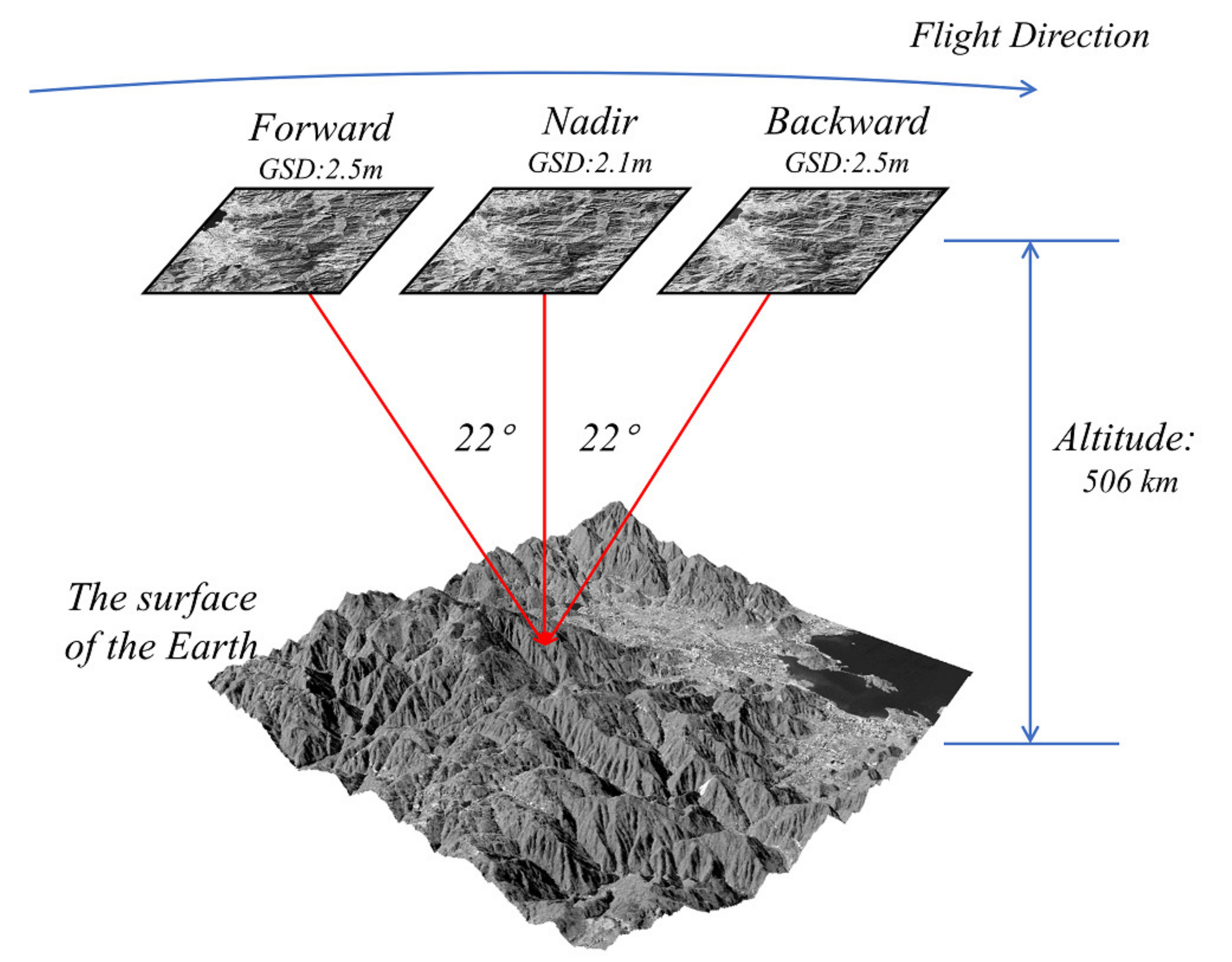}
\end{center}
\setlength{\abovecaptionskip}{-0.3cm}
   \caption{The three-line camera (TLC) of the ZY-3 satellite.}
\label{fig:4}
\vspace{-0.3cm}
\end{figure}
\begin{table*}
\small
\begin{center}
\renewcommand\arraystretch{1.15}
\setlength{\tabcolsep}{5 mm}{
\begin{tabular}{c|cccccc}
\hline
\textbf{Image size(pixels)}   & \textbf{768$^2$} & \textbf{4608$^2$} & \textbf{9216$^2$} & \textbf{13824$^2$} & \textbf{18432$^2$} & \textbf{23040$^2$} \\ \hline
Min fitting error(pixels)     & 0.00015          & 0.00010           & 0.00181           & 0.00152            & 0.00106            & 0.00257            \\ \hline
Max fitting error(pixels)     & 0.16204          & 1.09168           & 2.42714           & 3.89724            & 5.35514            & 6.62046            \\ \hline
\end{tabular}}
\end{center}
\setlength{\abovecaptionskip}{-0.2cm}
\setlength{\belowcaptionskip}{-0.2cm}
\caption{The error of fitting an RPC model with a pin-hole model increases with the size of the image patch.}
\label{tab:1}
\vspace{-0.3cm}
\end{table*}
\section{The Satellite MVS Dataset}
\subsection{Data Source}
This section describes the satellite MVS dataset we built, which is named the TLC SatMVS dataset. The triple-view images were collected from the TLC camera mounted on the Ziyuan-3 (ZY-3) satellite. The ground resolution of the nadir and the two side-looking images is 2.1 m and 2.5 m, respectively (see Fig.\! \ref{fig:4}). As a professional satellite for surveying and 3D mapping, the ZY-3 satellite accesses the same scene at almost the same time, without the impact of illumination and seasonal changes, which differs from the WorldView series with single linear-array cameras. The RPC parameters have been calibrated in advance to achieve a sub-pixel reprojection accuracy. The ground-truth DSMs were prepared from both high-accuracy LiDAR observations and ground control point (GCP)-supported photogrammetric software \cite{RN01}. The DSM was stored as a 5-m resolution regular grid under the WGS-84 geodetic coordinate system and the UTM projection coordinate system.
\subsection{The TLC SatMVS Dataset}
\begin{figure}
\begin{center}
\includegraphics[width=1.0\linewidth]{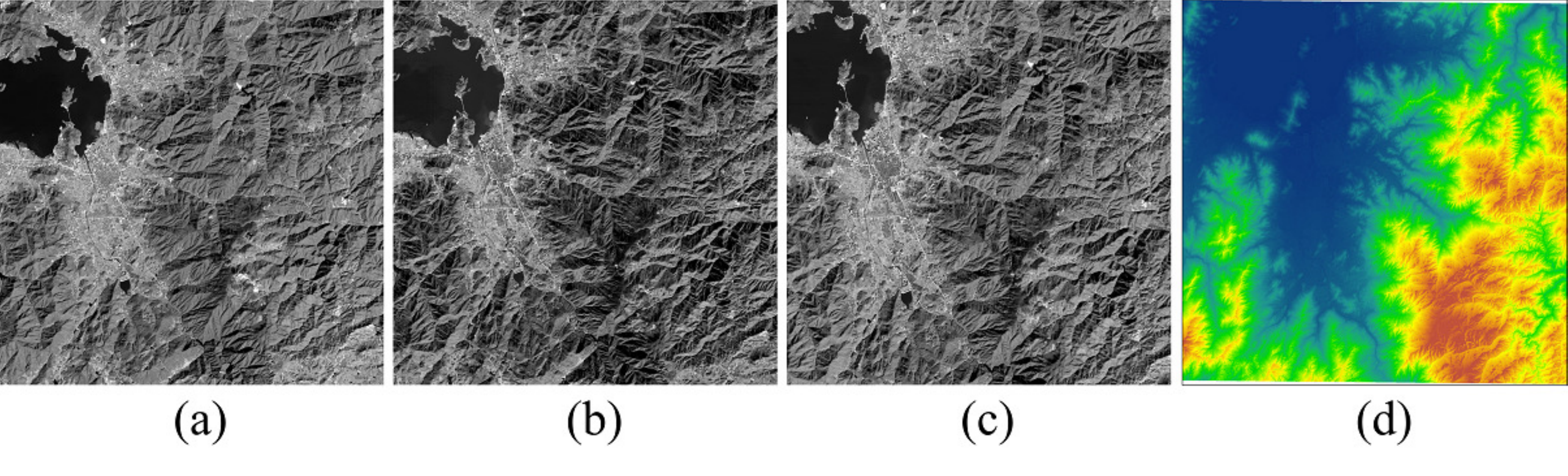}
\end{center}
\setlength{\abovecaptionskip}{-0.3cm}
   \caption{Examples of the collection of large-size satellite images ((a), (b), and (c) are the backward, forward, and nadir view images) and the corresponding ground truth (d) from the TLC SatMVS dataset.}
\label{fig:5}
\vspace{-0.3cm}
\end{figure}

\begin{figure}
\begin{center}
\includegraphics[width=1.0\linewidth]{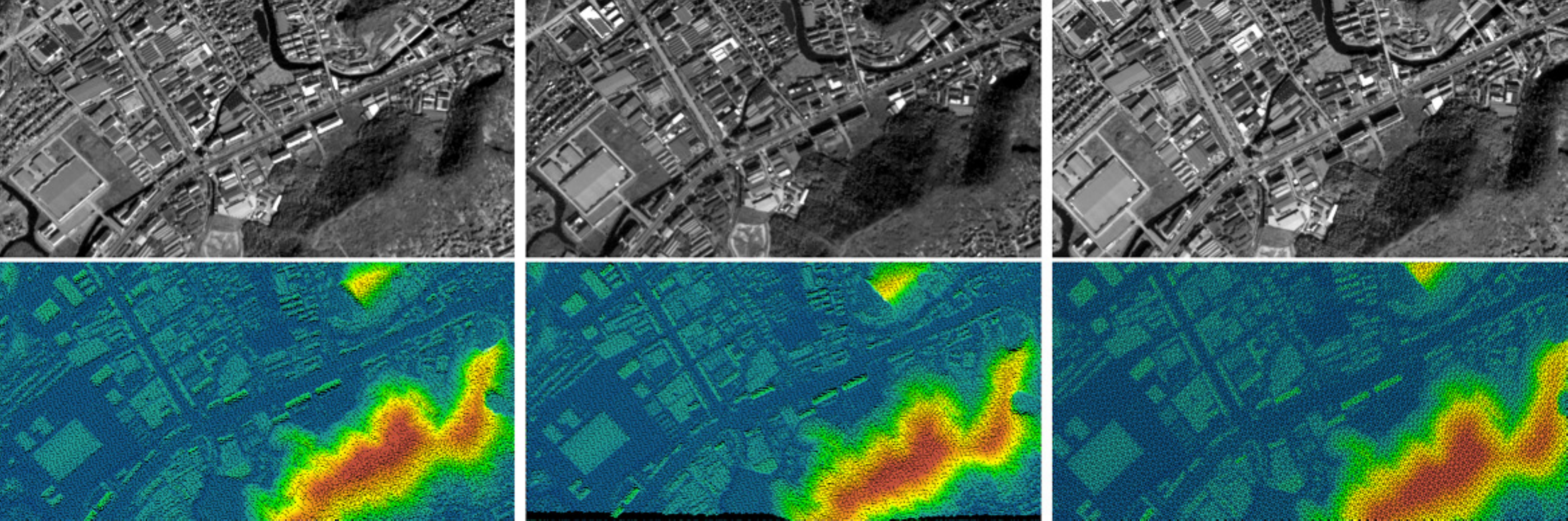}
\end{center}
\setlength{\abovecaptionskip}{-0.2cm}
   \caption{The ready-made version of the TLC SatMVS dataset for the training. The images from left to right are the image patches of the backward, forward, and nadir views, with the corresponding height maps below.}
\label{fig:6}
\vspace{-0.4cm}
\end{figure}

We built two versions of the TLC SatMVS dataset. The first version is a collection of large-size satellite images, and the second is a ready-made version for the training and testing of a learning method with mainstream GPU capacity.\\
In the first version (see Fig.\! \ref{fig:5}), there are 173 sets of images (we call one triple-view image a set), with 127 sets separated for training and the rest for testing. Each set contains 16-bit panchromatic triple-view images with a $5120\times5120$ pixel size, the RPC parameters, and the ground-truth DSM, covering approximately 125 $km^2$. The overlap rate of the triple-view images is more than 95\%.\\
The second version (ready-made version) contains cropped patches of the images (see Fig.\! \ref{fig:6}), the RPC parameters of each patch, and the corresponding height maps, which were obtained by projecting the DSMs to the images with the RPC parameters. The height map is theoretically equivalent to the depth map in a close-range MVS dataset, but stores the height instead of the depth information of the corresponding pixel in the image. Specifically, each $5120\times5120$ image is cropped into $768\times384$ patches with an overlap rate of 5\% in both the horizontal and vertical direction. There are 5011 sets for training in total. \\
We also provide an accessary version, where we fitted each patch from the RPC projection into pin-hole projection according to \cite{Zhang2019LeveragingVR} under the UTM coordinate system. The depth maps were obtained by projecting the ground truth to the patches with the fitted pin-hole camera parameters.  
\section{Experiments}
\subsection{Fitting the RPC Model with the Pin-hole Model}
Before this work, the complex RPC model was fitted with a pin-hole model in the conventional satellite MVS \cite{Zhang2019LeveragingVR}. The fitting error is unavoidable, and increases with the size of the image patch. As shown in Table \ref{tab:1}, the maximum fitting error is small in the patch with a size of $768\times768$ pixels, but it can reach more than 6 pixels at a size of $23040\times23040$ pixels, which is approximately the size of a ZY-3 TLC image or a WorldView-3 image. Fig. \ref{fig:7} also shows the error distributions in the XY-planes. Note that there are no geometric errors in the rigorous RPC model.
\begin{table*}[]
\footnotesize
\begin{center}
\renewcommand\arraystretch{1.15}
\setlength{\tabcolsep}{4mm}{
\begin{tabular}{c|cccccc}
\hline
\textbf{Methods} & \textbf{MAM(m)} & \textbf{RMSM(m)} & \textbf{\textless{}2.5m(\%)} & \textbf{\textless{}7.5m(\%)} & \textbf{Comp.(\%)} & \textbf{Runtime} \\ \hline
adapted COLMAP\cite{Zhang2019LeveragingVR}                                                            & 2.227                                                         & 5.291                                                       & 73.35                                                                   & 96.00                                                                     & 79.10                                                         & 77min27s                                                                      \\ \hline
RED-Net\cite{Liu2020ANR}*                                                                  & 2.171                                                         & 4.514                                                       & 74.13                                                                   & 95.91                                                                     & 81.82                                                         & 9min15s                                                                       \\
CasMVSNet\cite{Gu2020CascadeCV}*                                                                & 2.031                                                         & 4.351                                                       & 77.39                                                                   & 96.53                                                                     & \textbf{82.33}                                                & 4min02s                                                                       \\
UCS-Net\cite{Cheng2020DeepSU}*                                                                   & 2.039                                                         & 4.084                                                       & 76.40                                                                   & 96.66                                                                     & 82.08                                                         & \textbf{3min47s}                                                              \\ \hline
SatMVS(RED-Net)                                                                   & \textbf{1.945}                                                & 4.070                                                       & \textbf{77.93}                                                          & 96.59                                                                     & 82.29                                                         & 13min52s                                                                      \\
SatMVS(CasMVSNet)                                                                 & 2.020                                                         & \textbf{3.841}                                              & 76.79                                                                   & \textbf{96.73}                                                            & 81.54                                                         & 12min20s                                                                      \\
SatMVS(UCS-Net)                                                                   & 2.026                                                         & 3.921                                                       & 77.01                                                                   & 96.54                                                                     & 82.21                                                         & 13min17s                                                                      \\ \hline
\end{tabular}}
\end{center}
\setlength{\abovecaptionskip}{-0.2cm}
\caption{Quantitative results of the different MVS methods on the TLC SatMVS dataset. The proposed SatMVS with RPC warping implements three different learning-based MVS methods for height inference. Recent deep MVS methods imbedded with fitted homography warping are marked with *.)}
\label{tab:2}
\vspace{-0.3cm}
\end{table*}
\begin{table*}
\footnotesize
\begin{center}
\renewcommand\arraystretch{1.15}
\setlength{\tabcolsep}{3mm}{
\begin{tabular}{c|cccccc}
\hline
\textbf{Methods}                                                       & \textbf{MAE(m)}                                          & \textbf{RMSE(m)}                                         & \textbf{\textless{}2.5m(\%)}                            & \textbf{\textless{}7.5m(\%)}                            & \textbf{Comp.(\%)}                                      & \textbf{Runtime} \\ \hline
\begin{tabular}[c]{@{}c@{}}RED-Net \\ ($2048\times1472$)\end{tabular}         & 2.171                                                    & 4.515                                                    & 74.13                                                   & 95.91                                                   & 81.82                                                   & 9min10s          \\ \hline
\begin{tabular}[c]{@{}c@{}}RED-Net \\ ($5120\times5120$)\end{tabular}         & \begin{tabular}[c]{@{}c@{}}2.517\\ \textcolor{red}{(+0.346)}\end{tabular} & \begin{tabular}[c]{@{}c@{}}4.873\\ \textcolor{red}{(+0.358)}\end{tabular} & \begin{tabular}[c]{@{}c@{}}66.42\\ \textcolor{red}{(-7.71)}\end{tabular} & \begin{tabular}[c]{@{}c@{}}95.53\\ \textcolor{red}{(-0.38)}\end{tabular} & \begin{tabular}[c]{@{}c@{}}81.44\\ \textcolor{red}{(-0.38)}\end{tabular} & 4min17s          \\ \hline
\begin{tabular}[c]{@{}c@{}}SatMVS(RED-Net) \\ ($2048\times1472$)\end{tabular} & 1.945                                                    & 4.071                                                    & 77.93                                                   & 96.59                                                   & 82.29                                                   & 13min12s         \\ \hline
\begin{tabular}[c]{@{}c@{}}SatMVS(RED-Net) \\ ($5120\times5120$)\end{tabular} & \begin{tabular}[c]{@{}c@{}}1.946\\ \textcolor{red}{(+0.001)}\end{tabular} & \begin{tabular}[c]{@{}c@{}}4.224\\ \textcolor{red}{(+0.153)}\end{tabular} & \begin{tabular}[c]{@{}c@{}}77.88\\ \textcolor{red}{(-0.05)}\end{tabular} & \begin{tabular}[c]{@{}c@{}}96.54\\ \textcolor{red}{(-0.05)}\end{tabular} & \begin{tabular}[c]{@{}c@{}}82.35\\ \textcolor{red}{(+0.06)}\end{tabular} & 5min52s          \\ \hline
\end{tabular}}
\end{center}
\setlength{\abovecaptionskip}{-0.2cm}
\caption{Quantitative results of the SatMVS(RED-Net) and the RED-Net(with fitted pinhole model) on the TLC SatMVS dataset with different sizes.}
\label{tab:3}
\vspace{-0.3cm}
\end{table*}
\subsection{Model Evaluation}
\textbf{Implementation Details.} We evaluated the proposed SatMVS framework with imbedded RPC warping on the TLC SatMVS dataset. The framework was implemented in PyTorch and trained on a single NVIDIA TITAN RTX GPU (24GB). Different MVS architectures, including RED-Net, CasMVSNet, and UCS-Net, were integrated into the proposed framework. The hyper-parameters followed the same settings in all of the experiments: in the training phase, the batch size was set to 1, and RMSprop was selected as the optimizer. All the networks were trained for 35 epochs with an initial learning rate of 0.001, and were downscaled by a factor of 2 after the 10th epoch.\\
Three-stage hierarchical matching was adopted to infer the coarse-to-fine height or depth maps. For TLC images, the view number of the input images $N$ is fixed to 3. The numbers of hypothetical height planes were set to \{64, 32, 8\} in the three stages, and the corresponding interval was \{$(d_{max}-d_{min})$/64, 5$m$, 2.5$m$\}, expect for the UCS-Net implementation, where its own adaptive interval determination strategy was applied.\\ 
We also performed experiments on the MVS pipeline, with the fitted homography warping module used to replace the RPC warping module. Please note that double-precious floating point is used here to get rid of the numerical precision issue. As for adapted COLMAP \cite{Zhang2019LeveragingVR}, since it is itself a complete conventional satellite MVS pipeline, we directly used its own framework for the reconstruction.\\
\begin{figure}
\begin{center}
\includegraphics[width=0.9\linewidth]{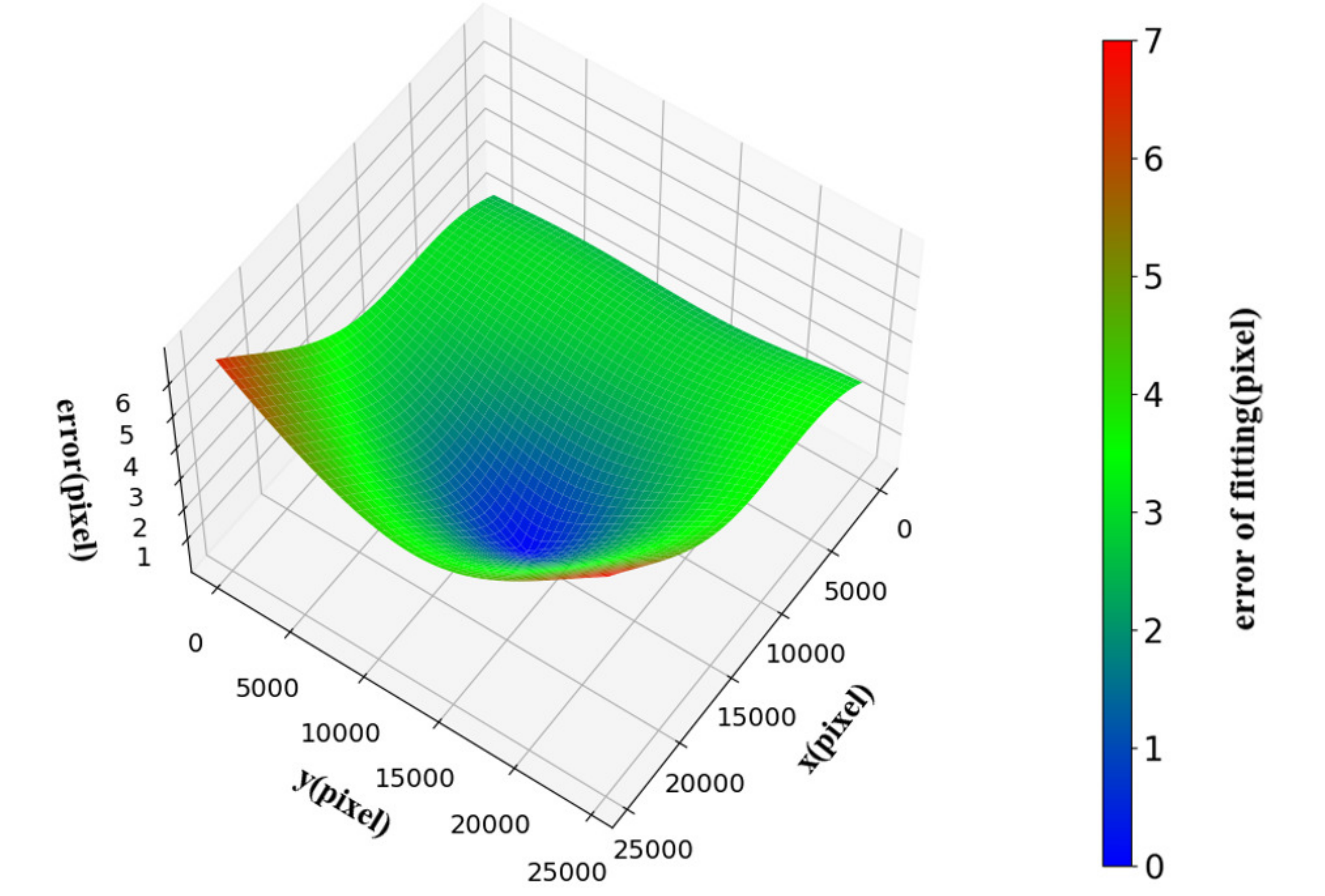}
\end{center}
\setlength{\abovecaptionskip}{-0.2cm}
   \caption{Error distribution of the RPC-to-pinhole model fitting. The X and Y coordinates represent the column and row coordinates of the image patches, and the Z axis represents the fitting error.}
\label{fig:7}
\vspace{-0.7cm}
\end{figure}
We adopt four commonly used metrics to evaluate the quality of the final DSM: 1) the mean absolute error (\textbf{MAE}), i.e., the average of the L1 distance over all the grid units between the ground truth and the estimated DSM; 2) the root-mean-square error (\textbf{RMSE}), i.e., the standard deviation of the residuals between the ground truth and the estimation; 3) the percentage of grid units with an L1 distance error below the thresholds of 2.5 m (approximately equivariant to the ground sample distance (GSD)) and 7.5 m ($<$\textbf{2.5m} and $<$\textbf{7.5m}); and 4) the completeness (\textbf{Comp.}), i.e., the percentage of grid units with valid height values in the final DSM.\\
\begin{figure*}
\begin{center}
\includegraphics[width=0.8\linewidth]{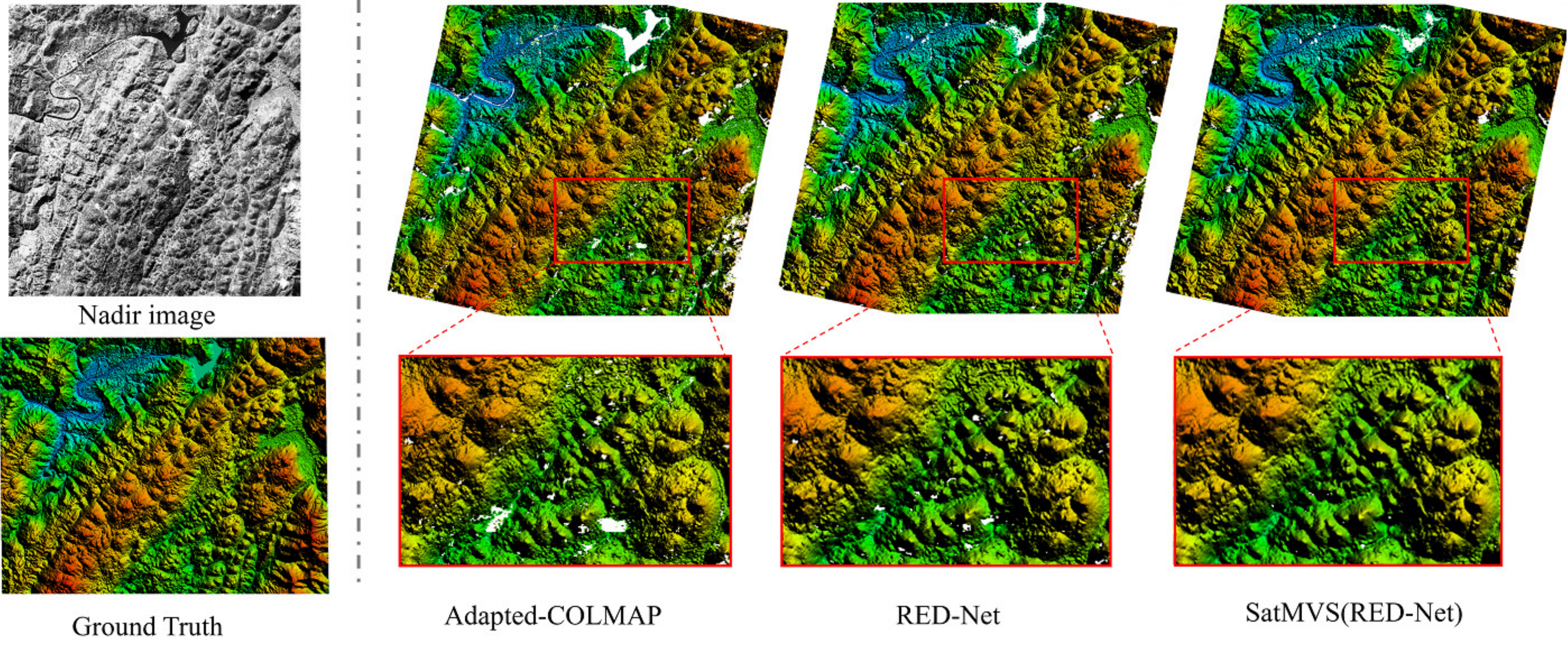}
\end{center}
\setlength{\abovecaptionskip}{-0.5cm}
   \caption{Results of RED-Net(homo), RED-Net(RPC), and adapted COLMAP\cite{Zhang2019LeveragingVR}.}
\label{fig:8}
\vspace{-0.2cm}
\end{figure*}
\begin{figure*}
\begin{center}
\includegraphics[width=0.9\linewidth]{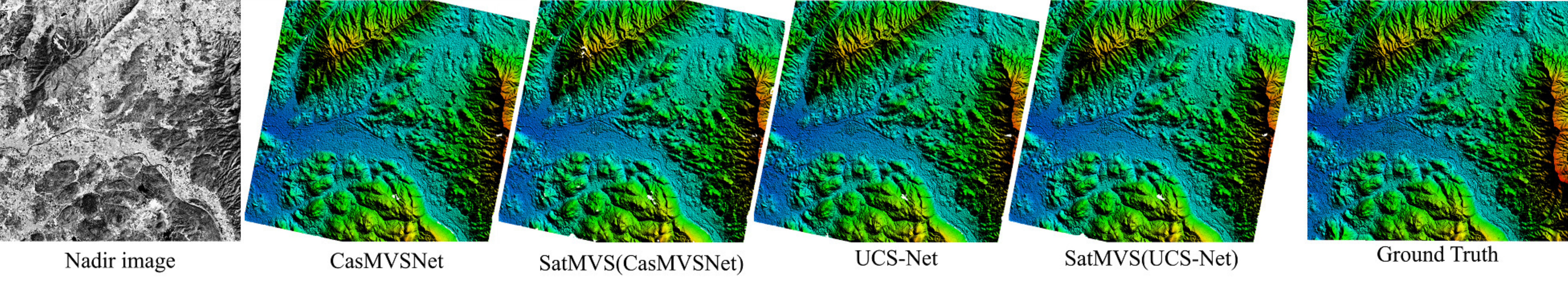}
\end{center}
\setlength{\abovecaptionskip}{-0.3cm}
   \caption{The qualitative results of CasMVSNet and UCS-Net with RPC warping and homography warping.}
\label{fig:9}
\vspace{-0.3cm}
\end{figure*}
\textbf{Evaluation on the TLC SatMVS Dataset.} As there are currently no deep learning based MVS methods for satellite images, we only compared the proposed SatMVS framework with a recent conventional method, i.e., adapted COLMAP \cite{Zhang2019LeveragingVR}, which uses the pin-hole camera model to fit the RPC model. In addition, we imbed the recent learning based methods [5,10,20] with the fitted homography warping into our pipeline to generate DSMs for comparison. In contrast, our SatMVS applies the rigorous RPC warping with different MVS regularization methods In the pipeline the images of the test set are cropped into patches with a size of approximately 2048$\times$1472 pixels for inference. The results are listed in Table \ref{tab:2}.\\
Several conclusions can be drawn from Table \ref{tab:2}. Firstly, all of the learning based MVS methods perform better than the conventional adapted COLMAP method. The advantage is apparent in the RMSE metric, which indicates that the learning methods have a lower variance of inference. Fig. \ref{fig:8} shows samples of the reconstructed DSMs produced by RED-Net(with fitted homography warping), SatMVS(RED-Net), and adapted COLMAP. The regions that failed to match are colored in white, most of which suffer from challenging scenes (occlusions, clouds, shadows, and textureless water surfaces). Compared with adapted COLMAP, the DSM results of the learning methods are more complete, especially the RPC warping versions.\\
Secondly, the performance of using the rigorous RPC model is only slightly better than using the fitted model. The reason is that the fitted model can also reach a sub-pixel accuracy at the size of 2048$\times$1472 pixels, as shown in Table \ref{tab:1}. \\
Thirdly, the inference of the deep learning based methods is much faster than that of the traditional adapted COLMAP method. However, the speed of RPC warping is lower than that of homography warping as the latter is only built on a simple $3\times3$ matrix multiplication operation.\\
\textbf{Evaluation on larger images.} We process the full-size images (5120$\times$5120 pixels) with RED-Net and our SatMVS(RED-Net) on the NVIDIA RTX A6000 GPU(48GB). The results are shown as Table 3. When processing 5120$\times$5120 images, it is observed that, (1) the performance of RED-Net severely decreases due to the increase of fitting error, but our RPC warping method remains stable; (2) SatMVS significantly outperforms RED-Net, which confirms the effectiveness and advantage of the proposed method; (3) the efficiency of both methods is significantly improved and reaches a comparable level. In addition to the high efficiency, using larger-capacity images have significantly simplified the process because cropping images into highly overlapped small patches is difficult when DSM is unavailable and processing the small patches will cause repeated and redundant matching in overlapped regions. It should be mentioned that here we only use RED-Net as the memory consumption of using CasMVSNet and UCS-Net is unaffordable.
\section{Discussion}
\vspace{-0.15cm}
We did not perform any more experiments on other datasets. The reason for this is that there are very few datasets that are suitable for the satellite MVS problem. The existing MVS3D dataset \cite{Bosch2016AMV} is very small, and was designed for testing non-learning based methods. The US3D dataset\cite{Bosch2019SemanticSF} is aimed at the joint reconstruction and segmentation task, where the stereo images of different views vary dramatically in illumination and seasonal changes. We failed to train an effective deep learning model with either of these datasets. The presented TLC SatMVS dataset greatly relieve this situation and will contribute to the study of deep learning based satellite MVS method. Furthermore, the TLC camera is more suitable for Earth surface reconstruction than the single linear array cameras such as those mounted on the WorldView series.
\vspace{-0.15cm}
\section{Conclusion}
In this paper, we have proposed the rigorous RPC warping model for the satellite MVS task for the first time. The advantage of the warping model is demonstrated both on the tests of the simulation data and the TLC SatMVS dataset. The experiments also show that the proposed learning based SatMVS framework performs better than the conventional method and the SOTA learning based methods when processing larger-capacity satellite images. In addition, we have presented the high-quality TLC SatMVS dataset. We believe that our work will promote the development of Earth surface reconstruction from MVS satellite images.
{\small
\bibliographystyle{ieee_fullname}
\bibliography{egpaper_for_review}
}
\end{document}